\documentclass[prl,aps,epsf,amsfonts,floats,amssymb,amsmath,groupedaddress,showpacs,floatfix,nofootinbib]{revtex4}

\usepackage{graphicx}
\usepackage[]{latexsym}
\usepackage{bm}

\usepackage{graphicx,amssymb,amsmath}
\usepackage[english]{babel}
\usepackage{graphics}                 
\usepackage{color}                    

\newcommand{\be}{\begin{equation}}\newcommand{\ee}{\end{equation}}
\newcommand{\bea}{\begin{eqnarray}}\newcommand{\eea}{\end{eqnarray}}
\newcommand{\beaa}{\begin{eqnarray}}\newcommand{\eeaa}{\end{eqnarray}}
\newcommand{\ba}{\begin{array}}\newcommand{\ea}{\end{array}}
\newcommand{\bit}{\begin{itemize}}\newcommand{\eit}{\end{itemize}}
\newcommand{\ben}{\begin{enumerate}}\newcommand{\een}{\end{enumerate}}

\def\lf{\left}

\def\non{\nonumber}

\def\ri{\right}

\def\1{{_{1}}}\def\2{{_{2}}}
\begin{document}

\title{Geometric phase and its applications to fundamental physics}

\author{ Antonio Capolupo}
\author{Giuseppe Vitiello}

 \affiliation{  Dipartimento di Fisica E.R.Caianiello and INFN gruppo collegato di Salerno,
  Universit\'a di Salerno, Fisciano (SA) - 84084, Italy}



\begin{abstract}

We report on recent results showing that the geometric phase can be used   as a tool in the analysis of many different physical systems, as mixed boson systems, CPT and CP violations,  Unruh effects,    and thermal states.
We show that the geometric phases appearing in the time evolution of mixed meson systems like $B_{s}^{0}-\bar{B}_{s}^{0}$ and the $K^{0}-\bar{K}^{0}$  are linked to the  parameter $z$ describing the $CPT$ violation.
A non zero phase difference between particle and antiparticle arises only  in the presence of  $CPT$ symmetry breaking.
 Then the geometric phase can represent a completely new test for the $CPT$ invariance.
Moreover, we study the geometric phase of systems represented by mixed state and undergoing a nonunitary evolution and  propose the realization of  interferometers which can prove the existence of the Unruh effect and can allow very precise measurements of temperature. We are glad to dedicate this paper to Professor Gaetano Vilasi in the occasion of his 70th birthday.

\end{abstract}

\maketitle

\section{Introduction}

In the recent years, many studies have been devoted  to the analysis of the geometric phase  \cite{Berry:1984jv}--\cite{Bruno:2011xa}. It characterizes  the evolution of many physical systems and has   been detected in different ways \cite{Tomita}-\cite{Pechal}.

On the other hand, it has been shown \cite{Capolupo:2013xza,Capolupo:2015ina} that in all the systems where the vacuum condensates are generated \cite{Takahashi:1974zn}-\cite{ Birrell}, the Aharonov--Anandan invariant \cite{Anandan:1990fq}  (AAI) and the geometric phase are produced
 in their evolution. This fact has suggested that phenomena like Unruh \cite{Unruh:1976db}, Hawking \cite{Hawking:1974sw} and Parker effects \cite{Parker:1968mv,Schrodinger}, characterized by the presence of a vacuum condensate and  very hard to be detected, togheter with some aspect of quantum field theory in curved space, could be analyzed by means of the geometric phase of atomic systems which simulate such phenomena \cite{Capolupo:2013xza,Capolupo:2015ina}.
It has been shown also that geometric phases and invariants can be used to test CPT
symmetry in meson systems like kaons and $B_{s}^{0}-\bar{B}_{s}^{0}$ system \cite{capolupo2011}, to prove the existence of postulated particles like the axions
\cite{CapolupoLamb2015}, to test SUSY violation in thermal states \cite{Capolupo:2013ica}, to reveal the Hawking \cite{Capolupo:2013xza}  and Unruh \cite{Capolupo:2013xza,Capolupo:2015ina,Hu-Yu,Ivette1} effect and to
build a quantum thermometer \cite{Capolupo:2013xza,Capolupo:2015ina,Ivette}.

In the present paper, we report the results obtained by studying the geometric phase for mixed meson systems \cite{capolupo2011} and for quantum open systems represented by atoms accelerated in an electromagnetic field and interacting with thermal states \cite{Capolupo:2015ina}.
We reveal the relation between the geometric phase appearing in the evolution of mixed mesons and the parameter denoting the $CPT$ violation \cite{capolupo2011}. We show that a non-trivial phase difference between particle and antiparticle indicates unequivocally the $CPT$ symmetry breaking in mixed bosons. We also show that the phases can be useful for the study of the $CP$ violation and we do a numerical analysis for $B_{s}^{0}-\bar{B}_{s}^{0}$ mesons \cite{capolupo2011}.

We then study the geometric phase for quantum open systems  \cite{Wang} and we show that a detectable difference of the geometric phases can be revealed between atoms which are accelerated in an arm of an interferometer and atoms which are inertial in the other branch.   Such a phase difference is due only to the Unruh effect \cite{Capolupo:2013xza,Capolupo:2015ina}.
Moreover, we show that the difference between geometric phases produced by atoms interacting with two different thermal states allows to determine the temperature of a sample once the temperature of the other one  is known  \cite{Capolupo:2013xza,Capolupo:2015ina}.

 The use of the phase defined  in \cite{Wang} allows to consider time intervals arbitrary small and very low transition frequencies, as well as spontaneous emission rates characterizing  fine and hyperfine atomic structures.
Indeed, in the short time intervals which we consider, the number of spontaneously emitted particle is negligible and the systems are quasi-stable.
These facts permits to improve the results obtained in previous works  studying different systems  \cite{Hu-Yu,Ivette1,Ivette}.
We consider  the structure of the atomic levels of $^{85}Rb$, $^{87}Rb$ and  $^{133}Cs$ which  improves the detection of Unruh effect and permits  very precise temperature measurements.

The structure of the paper is the following: in Section 2 we diagonalize the effective Hamiltonian of mixed meson systems by means of a complete biorthonormal set of  states; we compute the geometric phase  for mixed mesons and we show their links with   $CPT$ and $CP $ violations. A numerical analysis  for $B_{s}$  mesons  is also presented.
In Section 3  we analyze the geometric phase for a two level atom undergoing a non-unitary evolution and   study the possible applications to the detection of the Unruh effect and to build a very precise thermometer. Section 4 is devoted to the conclusions.

We are glad to dedicate this paper to Professor Gaetano Vilasi in the occasion of his 70th birthday.

\section{Meson mixing, CP and CPT violations and geometric phase}

The particle mixing phenomenon has been analyzed thoroughly   in the contexts of quantum mechanics (QM) \cite{Kabir}--\cite{fidecaro} and of quantum field theory (QFT) \cite{BCRV01}--\cite{Blasone:2002jv}, \cite{Blasone:2005ae} --\cite{Capolupo:2010ek}. Although  the QFT analysis discloses features which cannot be ignored (see for example Refs.\cite{CapolupoPLB2004} \cite{Capolupo:2006et}--\cite{Capolupo:2015oua}), nevertheless
 a correct phenomenological description of systems as $B^{0}-\bar{B}^{0}$ can be also dealt with  in the context of QM. Therefore, in the following we analyze the mixed bosons in the context of QM.

The state vector of mixed boson systems  as $K^{0}$, $B_{d}^{0}$, $B_{s}^{0}$ and $D^{0}$ can be represented   as
$
|\psi (t)\rangle \,=\, \psi_{M^{0}} (t) |M^{0} \rangle \,+\,\psi_{\bar{M}^{0}} (t) |\bar{M}^{0} \rangle \,
+\,\sum_{n} d_{n}(t)|n\rangle\,,
$
where $M^{0}$ is the meson state ($K^{0}$, $B_{d}^{0}$, $B_{s}^{0}$ or $D^{0}$); $\bar{M}^{0}$ is the corresponding antiparticle,   $|n\rangle$ are the decay products, $t $ is the proper time, $\psi_{M^{0}} (t)$, $\psi_{\bar{M}^{0}} (t)$ and $d_{n}(t)$ are time dependent functions.
At initial time $t=0$ one has
$
|\psi (0)\rangle \,=\, \psi_{M^{0}} (0) |M^{0} \rangle \,+\,\psi_{\bar{M}^{0}} (0) |\bar{M}^{0} \rangle \,.
$
 The time evolution of $|\psi (t)\rangle $ can be  described, in the space formed by $|M^{0} \rangle$ and $ |\bar{M}^{0} \rangle $  by means of the Schr\"odinger equation
$i \frac{d }{dt} \Psi   \,=\,\mathcal{H}\, \Psi  \,,$
where $\Psi = \lf(\psi_{M^{0}}(t)\,, \psi_{\bar{M}^{0}} (t) \ri)^{T}$ and $ \mathcal{H}$ is the effective  non-Hermitian Hamiltonian of the system, $ \mathcal{H} = \left(\begin{array}{cc}
                  \mathcal{H}_{11} & \mathcal{H}_{12} \\
                  \mathcal{H}_{21} & \mathcal{H}_{22} \\
                \end{array}\right)\,.$
 Here $\mathcal{H} = M - i \frac{\Gamma}{2}$, with $M$ and $\Gamma$ Hermitian matrices.
The matrix elements of $\mathcal{H}$ are constrained by the conservation of discrete symmetries  \cite{fidecaro}. Indeed,
$CPT$ conservation implies $\mathcal{H}_{11} = \mathcal{H}_{22}$, $T$ conservation requires $|\mathcal{H}_{12}| = |\mathcal{H}_{21}|$ and $CP$ conservation imposes
$\mathcal{H}_{11} = \mathcal{H}_{22}$ and   $|\mathcal{H}_{12}| = |\mathcal{H}_{21}|$.

In particular, the  $CP$ violation, (i.e. $|\mathcal{H}_{12}| \neq |\mathcal{H}_{21}|$), makes the Hamiltonian  $\mathcal{H}$  non-Hermitian, $\mathcal{H} \neq \mathcal{H}^{\dag}$ and non-normal,
$[\mathcal{H} , \mathcal{H}^{\dag}] \neq 0$\footnote{the matrices $M$ and $\Gamma$ do not commute,  $[M , \Gamma] \neq 0$}.
Thus, the left and right eigenstates of  $\mathcal{H}$ are independent sets of vectors
that are not connected by complex conjugation, then the diagonalization of  $\mathcal{H}$ needs the use  of non-Hermitian quantum mechanics. We will use the biorthonormal basis formalism \cite{Bender98} - \cite{Dattoli}.

We denote with $\lambda_{j}  = m_{j} - i \Gamma_{j}/2$, with $j=L,H$ ($L$ denotes the light mass state and $H$ the heavy mass state) the eigenvalues of the Hamiltonian $\mathcal{H}$ and
with $|M_{j}\rangle $,  the corresponding eigenvectors,
$
\mathcal{H} |M_{j}\rangle\, =\,  \lambda_{j}  |M_{j}\rangle \,.
$
Denoting with $|\widetilde{M}_{j}\rangle $, $(j=L,H)$ the eigenvectors of $\mathcal{H}^{\dag}$, the eigenvalues of $\mathcal{H}^{\dag}$ are the complex conjugate of those of
$\mathcal{H} $,
$\label{autovaHdag}
\mathcal{H}^{\dag}|\widetilde{M}_{j} \rangle  \, =\,  \lambda^{*}_{j} |\widetilde{M}_{j}\rangle  \,.
$
Notice that
the conjugate states $\langle\widetilde{M}_{j}|^{\dag} \equiv |\widetilde{M}_{j}\rangle$ and $| {M}_{j}\rangle^{\dag} \equiv \langle {M}_{j}|$  are not isomorphic to their duals:   $|\widetilde{M}_{j}\rangle \neq |{M}_{j}\rangle$
   and
 $\langle {M}_{j}| \neq \langle\widetilde{M}_{j} |$.
 In this case, a complete biorthonormal set for $\mathcal{H}$ is given by $ \{|M_{j}\rangle ,  \langle\widetilde{M}_{j}| \}$, with $j=L,H$. Indeed one has  the following biorthogonality relation
$
\langle\widetilde{M}_{i} |M_{j}\rangle\,= \langle M_{j}|\widetilde{M}_{i}\rangle\,=\,\delta_{ij} \,,
$
and  the completeness relations
$
\sum_{j } |M_{j}\rangle \langle\widetilde{M}_{j}| =\sum_{j } |\widetilde{M}_{j}\rangle \langle M_{j}| =1\,.
$
The existence of a complete biorthonormal set of eigenvector of  $\mathcal{H}$  implies that $\mathcal{H}$ is diagonalizable.

Moreover, since the time evolution operator associated with $\mathcal{H}$, $U(t)= e^{-i \mathcal{H} t}$  is not unitary, then we introduce the time evolution operator of $\mathcal{H}^{\dag}$, $\widetilde{U}(t)= e^{-i \mathcal{H}^{\dag} t}$, such that $U \widetilde{U}^{\dag} =
\widetilde{U}^{\dag} U = 1$.
%
The time evolved of the states $|M_{k} \rangle$ and $|\widetilde{M}_{k} \rangle$, $(k=L,H)$  are thus
$|M_{k}(t)\rangle = U(t) |M_{k}\rangle = e^{- i \lambda_{k}t}|M_{k} \rangle $ and $|\widetilde{M}_{k}(t)\rangle = \widetilde{U}(t) |\widetilde{M}_{k} \rangle = e^{- i \lambda^{*}_{k}t} |\widetilde{M}_{k} \rangle$.
%
%
%
%
%

 By  introducing the $CP$ parameter
$
\varepsilon 
=\frac{|p/q|-|q/p|}{|p/q|+|q/p|}= \frac{|\mathcal{H}_{12}|-|\mathcal{H}_{21}|}{|\mathcal{H}_{12}|+|\mathcal{H}_{21}|}\,,
$
where
$
q/p\,=
\,\sqrt{\mathcal{H}_{21}/\mathcal{H}_{12} }\,,$ and the $CPT$ parameter,
 $
 z\,
 =\,\frac{(\mathcal{H}_{22}-\mathcal{H}_{11})}{\lambda_{L} - \lambda_{H}}\,,
 $
 we derive  the correct meson states $|M^{0}(t)\rangle$ and $|\bar{M}^{0}(t)\rangle$ in terms of the eigenstates of $\mathcal{H}$ $|M_{L}\rangle$ and
$|M_{H}\rangle$ \cite{capolupo2011}, which have to be used in the computations,
\bea\label{B0states1}
|M^{0}(t)\rangle &=& \frac{1}{2 p}\lf[\sqrt{1-z}\, |M_{L}\rangle\, e^{-i \lambda_{L}t}\,+\,
\sqrt{1+z}\, |M_{H}\rangle\, e^{-i \lambda_{H}t}\ri]\,,
\\\label{B0states2}
|\bar{M}^{0}(t)\rangle &=& \frac{1}{2 q}\lf[\sqrt{1+z}\, |M_{L}\rangle\, e^{-i \lambda_{L}t}\,-\,
\sqrt{1-z}\, |M_{H}\rangle\, e^{-i \lambda_{H}t}\ri]\,,
\\\label{B0states3}
\langle \widetilde{M^{0}}(t)| &=& p\,\lf[ \sqrt{1-z}\, \langle \widetilde{M_{L} }| \, e^{ i \lambda_{L}t}\,+\,
\sqrt{1+z}\, \langle \widetilde{M_{H} }|\, e^{i \lambda_{H}t}\ri]\,,
\\\label{B0states4}
\langle \widetilde{\bar{M}^{0}}(t)| &=& q\,\lf[ \sqrt{1+z}\, \langle \widetilde{M_{L} }| \, e^{ i \lambda_{L}t}\,-\,
\sqrt{1-z}\, \langle \widetilde{M_{H} }|\, e^{i \lambda_{H}t}\ri]\,.
\eea


We now study the  geometric phase for mixed mesons. We  consider the phase for pure states with a diagonalizable non-Hermitian Hamiltonian $\mathcal{H}_{NH}(t)$  \cite{Mukunda} and analyze its extension to the biorthonormal basis formalism.
%
In this case, the geometric phase is defined as \cite{capolupo2011}
$
\Phi _{NH}(t) =  \arg \langle \widetilde{\psi}_{NH}(0 )| \psi_{NH}(t )\rangle\ -  \Im \int_{0}^{t} \langle\widetilde{\psi}_{NH}(t^{\prime})| \dot{\psi}_{NH}(t^{\prime}) \rangle dt^{\prime}\,,
$
which is reparametrization invariant and it is invariant under  the complex gauge transformations \cite{capolupo2011}. Here
 $|\psi_{NH}(t)\rangle$ and $|\widetilde{\psi}_{NH}(t)\rangle $
are the solution to the Schr\"odinger equation $i(d/dt)|\psi_{NH}(t)\rangle = \mathcal{H}_{NH}(t) |\psi_{NH}(t)\rangle$ and to its adjoint equation   $i(d/dt)|\widetilde{\psi}_{NH}(t)\rangle = \mathcal{H}_{NH}^{\dag}(t) |\widetilde{\psi}_{NH}(t)\rangle$,  respectively.

In the particular case  of  mixed meson systems $M^{0}-\bar{M}^{0}$ one has the  following phases \cite{capolupo2011}:
\bea\label{FMM}
\Phi_{ M^{0} M^{0}}(  t) & = & \arg \langle \widetilde{M^{0}}(0 )| M^{0}(t )\rangle\, - \Im \int_{0}^{t}  \langle \widetilde{M^{0}}(t^{\prime})| \dot{M}^{0}(t^{\prime})\rangle  dt^{\prime}\,,
\\\label{FMbMb}
\Phi _{\bar{M}^{0}\bar{M}^{0}}(  t) & = & \arg \langle \widetilde{\bar{M}^{0}}(0 )| \bar{M}^{0}(t )\rangle\, - \Im \int_{0}^{t}  \langle \widetilde{\bar{M}^{0}}(t^{\prime})| \dot{\bar{M}}^{0}(t^{\prime})\rangle  dt^{\prime}\,,
\eea
  \bea\label{FM-Mb}
\Phi_{ M^{0} \bar{M}^{0}}(  t) & = & \arg \langle \widetilde{M^{0}}(0 )| \bar{M}^{0}(t )\rangle\, - \Im \int_{0}^{t}  \langle\widetilde{M^{0}}(t^{\prime})| \dot{\bar{M}}^{0}(t^{\prime})\rangle  dt^{\prime}\,,
\\\label{FMb-M}
\Phi _{\bar{M}^{0}M^{0}}(  t) & = & \arg \langle \widetilde{\bar{M}^{0}}(0 )| M^{0}(t )\rangle\, - \Im \int_{0}^{t}  \langle\widetilde{\bar{M}^{0}}(t^{\prime})| \dot{M}^{0}(t^{\prime})\rangle  dt^{\prime}\,.
\eea
$\Phi_{ M^{0} M^{0}}(  t)$ and $\Phi_{ \bar{M}^{0} \bar{M}^{0}}(  t)$ are the phases of the particle $M^{0}$ and of the antiparticle $\bar{M}^{0}$, respectively. They are  connected to $ CPT$ violation parameter. $\Phi_{ M^{0} \bar{M}^{0}}(  t) $ and $\Phi _{\bar{M}^{0}M^{0}}(  t) $ are the phases due to particle-antiparticle oscillations which are linked to $CP$ violation (see below).

Denoting with
%
$m = m_{L} + m_{H}$, $\Delta m = m_{H}-m_{L}$ and $\Delta \Gamma = \Gamma_{H}- \Gamma_{L}$,
and assuming $\frac{\Delta \Gamma t}{2} $ small,
 which is valid in the range $|\Delta t|< 15 ps$ used in the experimental analysis on $B^{0}-\bar{B}^{0}$ system \cite{Aubert:2004xga}--\cite{Aubert:2007bp},
    Eqs.(\ref{FMM}) and (\ref{FMbMb}) become
\bea\label{mucundB2}\non
\Phi _{M^{0} M^{0}}(  t) & \simeq &
\arg \lf[\cos\lf(\frac{\Delta m t}{2}\ri) + (\Im z - i \Re z )\sin \lf(\frac{\Delta m t}{2}\ri) \ri]
\, + \,
\frac{t }{2} \lf(\Delta m\,\Re z\,+\,\frac{ \Delta \Gamma}{2}\,\Im z \ri)\,,
\\
\label{mucundBbar2}\non
\Phi_{\bar{M}^{0} \bar{M}^{0}}(  t) & \simeq &
\arg  \lf[\cos\lf(\frac{\Delta m t}{2}\ri) - (\Im z - i \Re z )\sin \lf(\frac{\Delta m t}{2}\ri) \ri]
\,- \,
\frac{t }{2} \lf( \Delta m\,\Re z\,+\,\frac{ \Delta \Gamma}{2}\,\Im z \ri)\,,
\eea
respectively (the explicit form  of $\Phi _{M^{0} M^{0}}$ and $\Phi_{\bar{M}^{0} \bar{M}^{0}}$ for the general case are reported in Ref.\cite{capolupo2011}).
Notice that $\Phi _{M^{0} M^{0}}$ and $\Phi_{\bar{M}^{0} \bar{M}^{0}}$ depend on the real and imaginary part of the $CPT$ parameter $z$ and the difference phase, $\Delta \Phi (t) = \Phi _{M^{0} M^{0}}(  t) -\Phi_{\bar{M}^{0} \bar{M}^{0}}(  t) $ is due only to terms related to $z$. Indeed it is non--zero only in the presence of $CPT$ violation.
In the case of $CPT$ invariance, $z=0$,  one has
$ \label{equality}
\Phi^{CPT} _{M^{0} M^{0}}(  t) \, = \,\Phi^{CPT}_{\bar{M}^{0} \bar{M}^{0}}(  t)\,= \,\arg \lf[ \cos\lf(\frac{\Delta m t}{2}\ri)\ri]\,,
$
which is trivially equal to $0$ or $\pi$ and $\Delta \Phi(t) = 0$.

On the other hand, the phases $\Phi_{M^{0} \bar{M}^{0}}(t)$ and $\Phi _{ \bar{M}^{0}  M^{0}}(  t)$,
for $\frac{\Delta \Gamma t}{2} \ll 1$  and omitting second order terms in $z$, (see Ref.\cite{capolupo2011} for the general expressions)  are
\bea\label{mucundBBbar2}\non
\Phi_{M^{0} \bar{M}^{0}}(t) & \simeq & \frac{\pi}{2}\,-\,\frac{m t}{2}\,+\,\arg \lf[\frac{p}{q}  \sin \lf(    \frac{\Delta m t}{2}\ri)\ri]\,-\,\frac{\Delta m t}{2}\,  \Re \lf(\frac{p}{q} \ri)\,-\,\frac{\Delta \Gamma t}{2} \, \Im \lf(\frac{p}{q} \ri) \,,
\\
\label{mucundBbarB2}\non
\Phi_{\bar{M}^{0} M^{0} }(t) & \simeq & \frac{\pi}{2}\,-\,\frac{m t}{2}\,+\,\arg \lf[\frac{q}{p}  \sin \lf(    \frac{\Delta m t}{2}\ri)\ri]\,-\,\frac{\Delta m t}{2} \, \Re \lf(\frac{q}{p} \ri)\,-\,\frac{\Delta \Gamma t}{2} \, \Im \lf(\frac{q}{p} \ri)\,.
\eea
The phase difference is $\Phi_{M^{0} \bar{M}^{0}}(t) - \Phi_{\bar{M}^{0} M^{0} }(t) \neq 0$.
On the contrary, in  the case of $CP$ conservation one has
\bea\label{mucundCP}
\Phi^{CP}_{M^{0} \bar{M}^{0}}(t)\,=\,\Phi^{CP}_{\bar{M}^{0} M^{0} }(t)\,=\,\frac{\pi}{2}\,-\,(m  + \Delta m)\frac{ t}{2}\,+\,\arg \lf[\sin \lf(\frac{\Delta m t}{2}\ri)\ri]\,,
\eea
and there is no phase difference.

\emph{Numerical analysis:}
We present a numerical analysis of the phase related to $z$, $\Delta \Phi = \Phi_{M^{0} M^{0}} - \Phi_{\bar{M}^{0} \bar{M}^{0}}$  for the $B_{s}$   system.
Such a system is particulary appropriate for the study of non-cyclic phases since many particle oscillations occur within its  lifetime.
We use the following experimental data:
  $m_{s} = 1.63007 \times 10^{13} ps^{-1}$,
$\Delta m_{s} = 17.77 ps^{-1}$,
$ \Gamma_{s} = 0.678 ps^{-1}$,
$\Delta \Gamma_{s}  = -0.062 ps^{-1}$.
Moreover,  we assume, $-0.1\leq \Re z \leq 0.1$ and $-0.1\leq \Im z \leq 0.1 $  \cite{Aubert:2007bp}. Notice that, in such intervals for  $\Re z$ and $\Im z$,
the phase  $\Delta \Phi$ is weakly depending  on the value of $\Im z $,
so that one can fix an arbitrary value of $\Im z $ in the values interval  $[-0.1, 0.1]$ and study the non-cyclic phases as functions of  time for different values  of $\Re z   $.
 In Figs. $(1)$,  the phase is drawn   for $\Im z = 0 $. In order to  better show the behavior of the phases, the figures contain two plots A) and B)  of the same phase for sample values of $\Re z$ belonging to the intervals  $ [-0.1, 0]$ and $ [0, 0.1]$, respectively.
\begin{figure}
\begin{picture}(300,220)(0,0)
\put(-1,20){\resizebox{7.5 cm}{!}{\includegraphics{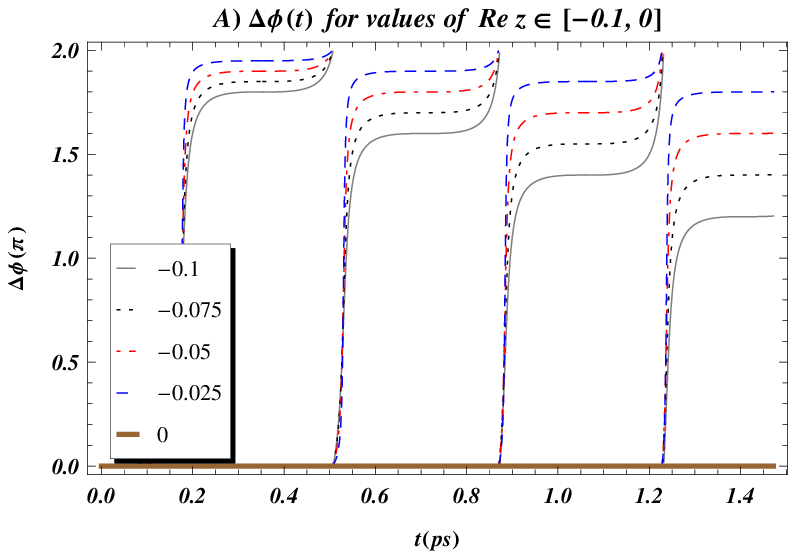}}}
\put(190,20){\resizebox{7.5 cm}{!}{\includegraphics{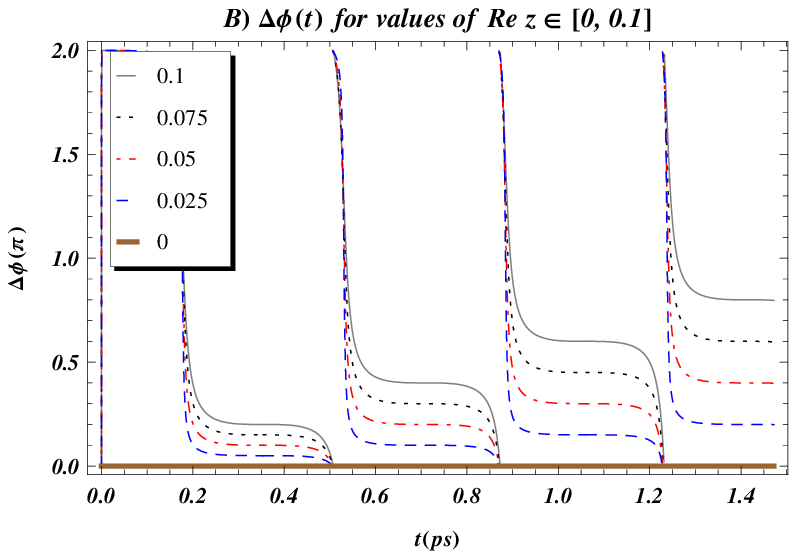}}}
\end{picture}
\caption{\em Plots of $\Delta \Phi= \Phi_{ B_{s}^{0} B_{s}^{0}} - \Phi_{ \bar{B}_{s}^{0} \bar{B}_{s}^{0}}$ as a function of time $t$ for   $\Im z = 0$ and different values of $\Re z$. In  picture A) $\Delta \Phi(t) $ is reported for sample values of $Re z \in [-0.1, 0]$ as indicated in the inset. In picture B) $\Delta \Phi(t) $ is plotted for sample values of $Re z \in [0, 0.1]$ as shown in the inset. }
\label{pdf}
\end{figure}
A  non--zero phase difference $\Delta \Phi$ appears in the case of $CPT$ violation.

\section{ Unruh effect, quantum thermometer and geometric phase}

In refs.\cite{Capolupo:2013xza,Capolupo:2015ina} we have shown that the phenomena characterized by the presence of vacuum condensate ~\cite{Unruh:1976db}-\cite{Birrell} are also characterized by the presence of Aharonov-Anandan invariant and of geometric phase.
Thus the geometric phase can used to study the properties of such systems and to reveal phenomena like Unruh, Casimir effects, or quantum field theory (QFT) in curved background, which are elusive to the observations.

Here   we focus our attention on the Unruh effect and on the possibility of the realization of a quantum thermometer. To do that we consider quantum open systems and use the Wang and Liu approach \cite{Wang} to define the geometric phase for mixed states in nonunitary, noncyclic evolution.

\emph{- Geometric phase for mixed states in nonunitary, noncyclic evolution:}

 The geometric phase is defined as
\bea\label{Wang2}
\Phi_{ g} & = &  \sum_{k=1}^{N} \arg \lf[\sqrt{ \lambda_{k}(t_{0})\lambda_{k}(t)} \,\langle
\varphi_{k}(t_{0})|\varphi_{k}(t )\rangle \ri]
\\\non
& - & \Im \sum_{k=1}^{N} \int_{t_{0}}^{t}   \lambda_{k}(t^{\prime}) \langle \varphi_{k}(t^{\prime})|\frac{\partial}{\partial t^{\prime}}|\varphi_{k}(t^{\prime}) \rangle
d t^{\prime}\,,
\eea
where $|\varphi_{k}(t ) \rangle$ and $\lambda_{k}(t)  $ are   eigenstates and  eigenvalues of the matrix density $\rho(t)$ describing the quantum open system for N-level mixed states.
The first term of Eq.(\ref{Wang2}) represents the total phase,  the second term the dynamic one.

In the case of a two level open system, $N =2$,
the radius of Block sphere is $r(t) =\sqrt{n_{1}^{2}+n_{2}^{2}+n_{3}^{2} }$, with
$n_{1} = \rho_{12} + \rho_{21} \,,
$
$n_{2} = i (\rho_{12} - \rho_{21}) \,,
$
$
n_{3} = \rho_{11} - \rho_{22}\, .
$
Defining the angles
$
\theta = \cos^{-1} \lf(n_{3}/r  \ri) $ and $  \phi = \tan \lf(n_{2}/n_{1}  \ri)\,,
$
the eigenvectors of $\rho$, $|\varphi_{1}(t )\rangle$ and $|\varphi_{2}(t )\rangle$  are (apart overall phase factors)
\bea\non
|\varphi_{1}(t )\rangle = \left(
                              \begin{array}{c}
                                 \cos \frac{\theta(t)}{2} \\
                                e^{i \phi(t)} \sin \frac{\theta(t)}{2} \\
                              \end{array}
                            \right)\,,
                            {}& &
\qquad
|\varphi_{2}(t )\rangle = \left(
                              \begin{array}{c}
                                \sin \frac{\theta(t)}{2} \\
                                - e^{i \phi(t)}\cos \frac{\theta(t)}{2} \\
                              \end{array}
                            \right)\,,
\eea
and the eigenvalues are $\lambda_{1}(t) = \frac{1}{2}[1 + r(t)]$ and $\lambda_{2} = \frac{1}{2}[1 - r(t)]$.
%

\emph{- Interaction:}
Let us now consider the Hamiltonian \cite{Compagno},
$
H = \frac{\hbar}{2}\,\omega_{0}\,\sigma_{3}\,+\,H_{F}\,-\,  \sum_{mn}\mathbf{\mu}_{mn}\cdot \mathbf{E}(x(t))\sigma_{mn}\,,
$
which describes a two level  atom as an open system with a non-unitary evolution in the reservoir of the electromagnetic field.
In $H$, $\omega_{0}$ is the energy level spacing of the atom, $\sigma_{3}$ is the Pauli matrix,  $H_{F}$ is the electromagnetic field Hamiltonian, $\mathbf{\mu}_{mn}$ is the matrix element of the dipole momentum operator connecting single-particle states $u_{n}$ and $u_{n^{\prime}}$ (\cite{Compagno}), $\sigma_{mn} = \sigma_{m }\sigma_{ n}$, and  $\mathbf{E}$ is the strength of the electric field.
 We consider a weak interaction between atom and field and study the evolution of the total density matrix $\rho_{tot}= \rho(0)\otimes |0\rangle \langle 0|$, in the frame of the atom.
 Here $|0\rangle$ and $\rho(0)$ are the vacuum and the initial reduced density matrix of the atom.
The evolution is given by \cite{Lind,Lind1}
 \bea\label{evolution}
\frac{\partial \rho(\tau)}{\partial \tau} & =& -\frac{i}{\hbar}[H_{eff},\rho(\tau)]
+ \frac{1}{2}\sum_{i,j=1}^{3} a_{i j } \lf(2 \sigma_{j} \,\rho\, \sigma_{i} - \sigma_{i}\, \sigma_{j}\, \rho - \rho \, \sigma_{i}\, \sigma_{j} \ri),
\eea
with  $\tau$  proper time,
$H_{eff}$  effective hamiltonian,
$
H_{eff}\,\simeq\,\frac{\hbar}{2}\,\omega_{0}\,\sigma_{3}\,,
$ ($\omega_{0}$ is the atomic transition frequency, we neglect the Lamb shift terms),
$a_{i j } $  coefficients of the Kossakowski matrix,
$
a_{i j } \,=\, \Sigma\, \delta_{i j } -i \Upsilon\, \epsilon_{i j k } \delta_{k 3} - \Sigma\, \delta_{i 3}\delta_{j 3},
$
with
$
\Sigma = \frac{1}{4}\lf[G(\omega_{0})+ G(-\omega_{0})\ri],
$
$
 \Upsilon = \frac{1}{4}\lf[G(\omega_{0})- G(-\omega_{0})\ri],
$
and
$
G(\omega)\,=\,\int_{-\infty}^{\infty} d \tau e^{i \omega \tau}G^{+}(x(\tau))\,
$
  Fourier transform of

$
  G^{+}( x-y)
  = \frac{e^{2}}{\hbar^{2}}\sum_{i,j=1}^{3} \langle +|r_{i}|-\rangle \langle -|r_{j}|+\rangle \langle 0 |E_{i}(x) E_{j}(x)|0 \rangle.
$

For an  initial state of the   atom given by
 $
|\psi(0)\rangle = \cos \lf(\frac{\theta}{2}\ri)|+\rangle + \sin \lf(\frac{\theta}{2}\ri)|-\rangle \,,
 $
with $\theta \equiv \theta (0)$, one has  the reduced density matrix $\rho(\tau)$ \cite{Capolupo:2013xza,Capolupo:2015ina,Hu-Yu}
\bea
\rho(\tau) =
\frac{1}{2}\left(
  \begin{array}{cc}
    \chi + 1  &  e^{  -i \Omega \tau}  \sqrt{\xi^{2} -\chi^{2}} \\
 e^{  i \Omega \tau}  \sqrt{\xi^{2} -\chi^{2}}  &  1- \chi \\
  \end{array}
\right)\,,
\eea
%
%
%
%
where
$
 \xi(\tau) = \sqrt{\chi^{2} + e^{-4 \Sigma   \tau}\sin^{2}\theta}\,,
 $ and
$
 \chi(\tau) = e^{-4 \Sigma \tau}\cos\theta + \frac{\Upsilon}{\Sigma}(e^{-4 \Sigma \tau}-1)\,.
 $
%

The eigenvalues and the corresponding eigenvectors of $\rho(\tau)$ are:
$
\lambda_{\pm} =  \frac{1}{2}(1 \pm \xi)\,,
$
and
\bea\non
|\phi_{+}(\tau)\rangle & = & \cos \lf(\frac{\theta(\tau)}{2}\ri)|+\rangle + \sin \lf(\frac{\theta(\tau)}{2}\ri) e^{i \Omega \tau}|-\rangle \,,
\\ {} &&
\\\non
|\phi_{-}(\tau)\rangle & = & \sin \lf(\frac{\theta(\tau)}{2}\ri)| +\rangle - \cos \lf(\frac{\theta(\tau)}{2}\ri) e^{i \Omega \tau}|-\rangle \,.
  \eea
Considering the initial time $t_{0} =0 $,
the geometric phase (\ref{Wang2})   becomes
\bea\label{fase}
\Phi_{ g}(t)   =   \arg \lf[\cos \frac{\theta}{2} \cos \frac{\theta(t)}{2}+
\sin \frac{\theta}{2} \sin \frac{\theta(t)}{2} e^{i \Omega t} \ri]
- \frac{\Omega}{2}\,\int_{0}^{t} \lf[ 1 - \xi(\tau) \cos \theta(\tau)\ri] d \tau\,,
\eea
with $\theta \equiv \theta (0)$. Such a phase can be used in the detection of Unruh effect and in the building of a quantum thermometer.

\emph{- Unruh effect:}

%

In the case of Unruh effect, we study the difference between  two geometric phase (\ref{fase}) computed for the two level system in the presence of an acceleration and in the inertial case. Such a phase difference is due only to the atom acceleration and then to the Unruh effect, since the accelerated system sees the Minkowski vacuum as a thermal Rindler vacuum.

Considering the atom acceleration through Minkowski spacetime  in the $x$ direction, the Rindler coordinates are
$x(\tau) = \frac{c^2}{a} \cosh \frac{a \tau}{c}$, $t(\tau) = \frac{c }{a} \sinh \frac{a \tau}{c}$.
Then the function  $\sin \frac{\theta(t)}{2} = \pm \sqrt{\frac{1}{2}\lf( 1 - \frac{\chi(t)}{\xi(t)} \ri)}$  in Eq.(\ref{fase}) becomes
\bea
\sin \frac{\theta_{a}(t)}{2} & = & \pm\, \sqrt{\frac{1}{2}-\frac{R_{a}\,-\,R_{a}\,e^{4 \Sigma_{a} t} + \cos \theta}{2 \sqrt{e^{4 \Sigma_{a} t}\sin^{2} \theta\,+\,\lf(R_{a}\,-\,R_{a}\,e^{4 \Sigma_{a} t}\,+\,\cos \theta \ri)^{2}}}} \,,
\eea
and similar for $\cos \frac{\theta_{a}(t)}{2} $, where $R_{a} = \Upsilon_{a}/\Sigma_{a}$, with
$
\Sigma_{a} = \frac{\gamma_{0}}{4} \,\lf(1+\frac{a^{2}}{c^{2} \omega_{0}^{2}}\ri)\,\frac{e^{2 \pi c \omega_{0}/a}+1}{ e^{2 \pi c \omega_{0}/a}-1 }
$ and
$
  \Upsilon_{a} = \frac{\gamma_{0}}{4} \,\lf(1+\frac{a^{2}}{c^{2} \omega_{0}^{2}}\ri),
$
and $\gamma_{0}\,$  spontaneous emission rate.

%

Eq.~(\ref{fase}) holds also for an inertial atom, $a=0$. In this case, $\sin \frac{\theta_{a}(t)}{2}$, $\cos \frac{\theta_{a}(t)}{2}$ and $\cos  \theta_{a}(t) $ are replaced by
 $\sin \frac{\theta_{a=0}(t)}{2}$, $\cos \frac{\theta_{a=0}(t)}{2}$ and $\cos  \theta_{a=0}(t) $ in which
 $\Sigma_{a}$, $\Upsilon_{a}$, $R_{a}$ are replaced by $\Sigma_{a=0}=\Upsilon_{a=0}=\gamma_{0}/{4}$,  $R_{a=0}=1$.

The phase difference  $\Delta \Phi_{U}(t) =   \Phi_{a}(t) - \Phi_{a=0}(t) $, gives the geometric phase in terms of the acceleration $a$.

Notice that the atomic element which has to be used in the interferometer plays an important role, since a non trivial value of $\Delta \Phi_{U}(t)$ can be achieved when $ \gamma_{0} / \omega_{0} > 10^{-5}$.
%
%

In Fig.2 we plot  $ \Delta \Phi_{U}$ as function of the acceleration $a$ for different hyperfine level structures of $^{85}Rb$, $^{87}Rb$ and $^{133}Cs$, as reported in the caption of the picture.
Values of  $\Delta \Phi_{U} \sim 10^{-4} \pi$, which are accessible with the current technology, can be obtained for  accelerations  of order of $10^{16} m/s^{2}$ and speeds of order of $(0.2-0.3) c$. The  time intervals considered are of order of $1/\omega_0$ in order to have negligible spontaneous emission ($N(t \sim 1/\omega_0) \sim 0.98 N(0)$).
\begin{figure}
\begin{picture}(300,180)(0,0)
\put(70,20){\resizebox{8.0 cm}{!}{\includegraphics{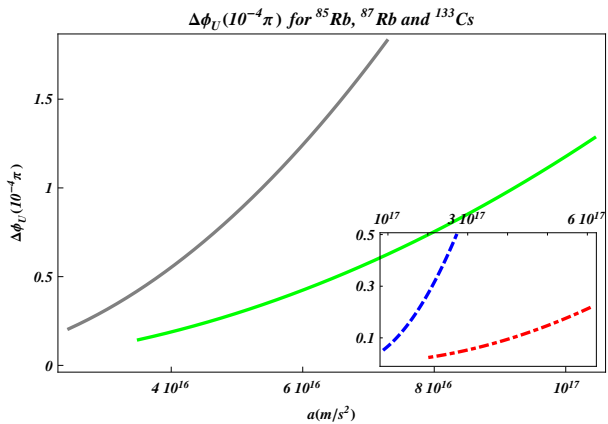}}}
\end{picture}\vspace{-1cm}
\caption{\em Plots of $\Delta \Phi_{U}$ as a function of the atom acceleration $a$, for  time intervals $t \simeq   1  /\omega_{0} $ and the splitting between the hyperfine energy levels:
main pictures: - (gray) dot dashed line: $^{87} Rb$, $5^{2}P_{1/2}$  line, splitting for $F=1 \rightarrow F=2 $ transition ($\omega_{0} = 814.5 MHz$, $\gamma_{0}= 36.129 MHz $ \cite{Daniel-Rb87}); - (green) solid line: $^{133}Cs$,   $6^{2}P_{1/2}$ line, splitting of the between  the $F=3$ and $F=4$ levels ($\omega_{0} = 1167.68 MHz$, $\gamma_{0}= 28.743 MHz $ \cite{Daniel-C}). Pictures in the inset:
- (blue) dashed line:  $^{85}Rb$, $5^{2}S_{1/2}$ line, energy splitting between the levels $F=1$ and $F=2$ ($\omega_{0} = 3.035 GHz$, $\gamma_{0}= 36.129 MHz $ for $D_{1}$ transition, $\gamma_{0}= 38.117 MHz $ for $D_{2}$ transition \cite{Daniel-Rb85}); - (red) dot dashed line:  $^{87}Rb$, $5^{2}S_{1/2}$ line, energy splitting between the levels $F=1$ and $F=2$ ($\omega_{0} = 6.843  GHz$, $\gamma_{0}= 36.129 MHz $ for $D_{1}$ transition, $\gamma_{0}= 38.117 MHz $ for $D_{2}$ transition \cite{Daniel-Rb87}).}
\label{pdf}
\end{figure}
%
%
%
%
The geometric phase above described can be revealed with a Mach-Zehnder interferometer with branches  of length of $4$ cm. In particular, a difference in the arm lengths of the interferometer of about  $0.1 \mu m$ permits to remove almost completely the dynamical phase differences $\delta $ and to reveal only the geometric phase \cite{Capolupo:2015ina}.

\emph{- Quantum thermometer:}

The geometric phase of Eq.(\ref{fase}) appears also when an atom interacts  with thermal states. Therefore, its analysis could allow very precise measurement of the temperature.

In the case of thermal states, the above coefficients $\Upsilon_{a}$ and $\Sigma_{a}$ are replaced by  $\Sigma_{T}$
 and $ \Upsilon_{T}$ which  depend on the temperature  \cite{Capolupo:2013xza,Capolupo:2015ina}, $\Upsilon_{T}\,=\,(\gamma_{0}/{4})\,(1+{4\pi^{2} k_{B}^{2}T^{2}}/{\hbar^{2} \omega_{0}^{2}})\,,$
 and
$\Sigma_{T}\,=\,(\gamma_{0}/{4})\,(1+{4\pi^{2} k_{B}^{2}T^{2}}/{\hbar^{2} \omega_{0}^{2}})\,({e^{E_{0}/k_{B} T}+1})/({e^{E_{0}/k_{B} T}-1})\,,$ $E_{0} = \hbar \omega_{0}\,.$

Then a quantum thermometer can be built by means of  an  interferometer in which an atom follows two different paths  interacting with two thermal states at different temperatures.
Assuming known the temperature of one thermal state, the temperature of the other one can be obtained by measuring the difference between the  geometric phases generated in the two paths. In the following we assume known the temperature $T_h$ of the hotter source and we derive the temperature $T_c$ of the colder source by measuring $\Delta \Phi_{T}$.

Also in this case we consider the hyperfine structure of $^{133} Cs$, and $^{87} Rb$  and we plot in Fig.~3, $\Delta \Phi_{T}$ as function of the temperatures of cold sources  $T_{c}$, for different  lines and different values of $T_{h}$.
 %
%
We consider time intervals  $t \simeq  \frac{1}{4 \omega_{0}} ~s $ in order that the  particle decay can be neglected
and we obtain temperatures of the cold source of $\sim 2$ orders of magnitude below the reference temperature of the hot source.
%
Paths of slightly different lengths can be chosen in order to let the geometric phase be dominating over the relative dynamical phase.

\begin{figure}
\begin{picture}(300,180)(0,0)
\put(70,20){\resizebox{8.0cm}{!}{\includegraphics{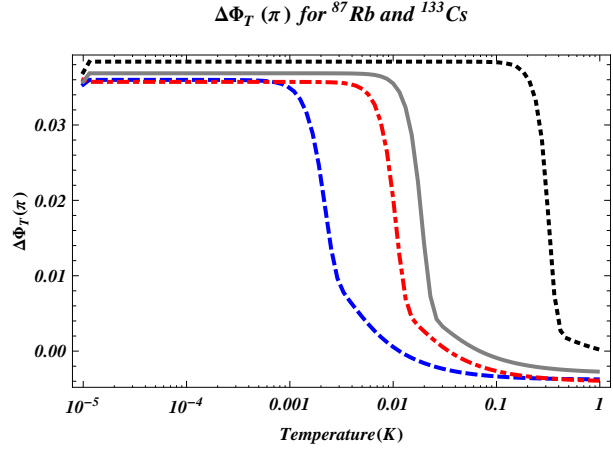}}}
\end{picture}\vspace{-1cm}
\caption{\em Plots of $\Delta \Phi_{T}$ as function of the temperatures of cold sources  $T_{c}$, for
 the splitting between the hyperfine energy levels and  $T_{h}$ values: - (blue) dashed line: $^{133} Cs$, $6^{2}P_{3/2}$ line, splitting for $F=4 \rightarrow F=5$ transition
  $(\omega_{0} = 251.09 MHz$, $\gamma_{0}= 32.889 MHz$ \cite{Daniel-C}) and $T_{h} = 10^{-2}K$; - (red)dot-dashed line: $^{87} Rb$, $5^{2}P_{1/2}$  line, splitting for $F=1 \rightarrow F=2 $ transition ($\omega_{0} = 814.5 MHz$, $\gamma_{0}= 36.129 MHz $ \cite{Daniel-Rb87}),   and $T_{h} = 3 \times 10^{-2}K$; - (gray) solid line: $^{133} Cs$, $6^{2}P_{1/2}$  line, splitting  for $F=3 \rightarrow F=4$ transition ($\omega_{0} = 1167.68 MHz$, $\gamma_{0}=  28.743 MHz $ \cite{Daniel-C}) and $T_{h} = 6 \times 10^{-2}K$; - (black) dotted line: $^{133} Cs$, $6^{2}S_{1/2}$ line, splitting  for $F=3 \rightarrow F=4$ transition ($\omega_{0} = 9.192 GHz$, $\gamma_{0}=  28.743 MHz $ \cite{Daniel-C}) and $T_{h} = 1 K$. The time considered are  $t \simeq   \frac{1}{4 \omega_{0}} ~s $. }
\label{pdf}
\end{figure}
%


\section{Conclusions }

We have studied the geometric phase for different phenomena and analyzed its possible applications.
We have shown that the geometric phases generated by the evolution of mixed  meson systems depend on the $CPT$ violating parameter  $z$ \cite{capolupo2011}.
In particular, we have shown that a non zero difference of phase $\Delta \Phi$ between particle and antiparticle  appears only in the case of $CPT$ symmetry breaking.
Therefore, in the next future, accurate analysis of the geometric phases for mixed mesons like the neutral $B_{s}$ system and  the kaons,
might represent a completely alternative method to test one of the most important symmetries of the nature. The geometric phase can allow also the study of  the $ CP$ symmetry breaking in mixed mesons.

Moreover, we have analyzed the geometric phase for mixed state with a non-unitary evolution and we have shown that the geometric phase of atoms accelerated in an interferometer  could permit the laboratory detection of the Unruh effect.
Similar atoms, interacting with two different thermal states can be utilized in an interferometer to have very precise temperature measurements \cite{Capolupo:2015ina}.

\acknowledgments
Partial financial support from MIUR and INFN is acknowledged.

\end{document}